\documentstyle[eqsecnum,aps]{revtex}

\begin{document}
\draft

\noindent \hspace*{12.8cm} (July 1996)\\
 
\begin{center}                                                                                                                                  
{\Large  \bf A Model-independent Way to Measure $|V_{ub}/V_{cb}|$\\}

\vspace{0.8cm}
 
C. S. Kim\footnote{Talk given  
at Asian Science Seminar (KOSEF-JSPS Winter School), 
Seoul Korea, on February 21 - 28 (1996).
Proceedings will be published by J. Kor. Phys. Soc. (1996).}\\

{\it Department of Physics, Yonsei University
Seoul 120-749, Korea}\footnote{kim@cskim.yonsei.ac.kr}\\
{\rm and}\\
{\it Theory Division, KEK, Tsukuba, 
Ibaraki 305, Japan}\footnote{cskim@kekvax.kek.jp}\\

\end{center}

\begin{abstract} 
We propose a new model-independent method, 
to determine the ratio $|V_{ub}/V_{cb}|$,
which is theoretically described by the phase space factor
and the well-known perturbative QCD correction only.
We explore the possible experimental options: the measurement of
inclusive hadronic invariant mass distributions,  
the `$D - \pi$' separation condition. 
We also discuss the relevant experimental backgrounds.
\end{abstract}

\vspace{0.5cm}
\baselineskip 22pt

\narrowtext

\noindent
{\bf 1. General Discussions}~~~
The CKM matrix element $V_{ub}$ is important to the SM description  
of CP-violation. If it were zero, there would be no CP-violation from 
the CKM matrix elements ({\it i.e.} in the SM), and we have to seek 
for other source of CP violation in $K_{L} \rightarrow \pi\pi$. 
Observations of semileptonic $b\rightarrow u$ transitions by the CLEO 
\cite{cleo} and ARGUS \cite{argus}  imply that $V_{ub}$ is  indeed nonzero, 
and it is important to extract the modulus $|V_{ub}|$ from semileptonic 
decays of $B$ mesons as accurately as possible.  
  
Presently, the charged lepton energy spectrum ($d \Gamma / d E_l$)
has been measured, and the $b\rightarrow u$ events are selected from the high 
end of the charged lepton energy spectrum.  This method is applied to both  
inclusive and exclusive semileptonic $B$ decays.  
However, this cut on $E_l$ is not very effective, since only  below 
$10 \%$ of $b\rightarrow u$ events survive this cut at the $B$ meson 
rest frame. (In the future asymmetric $B$-factories with boosted $B$ mesons, 
even much less than $10 \%$ would survive the $E_l$ cut over the 
$b \rightarrow c$ threshold.)
We also note that the dependences of  the lepton energy spectrum  on 
perturbative and non-perturbative QCD corrections \cite{kuhn,hqet} 
as well as on the unavoidable specific model parameters 
({\it e.g.} the parameter $p_{_F}$ of ACCMM model \cite{accmm}) are 
the strongest at the end point region, 
which makes the model-independent determination of $|V_{ub}/V_{cb}|$ 
almost impossible from the inclusive distribution of
$d \Gamma / d E_l$.
For exclusive $B \rightarrow X_u l \nu$ decays, the application of 
heavy quark effective theory (HQET) is much limited, since $u$-quark
is not heavy compared to $\Lambda_{QCD}$. And the theoretical predictions
for the required hadronic matrix elements are  largely
different depending on which model we use, as can be seen in the following,
as an example, for $\bar B^0 \rightarrow \rho^+ l^- \bar\nu$,
\begin{eqnarray}
\gamma_{\rho} \equiv 
{\Gamma_{theory}(\bar B^0 \rightarrow \rho^+ l^- \bar\nu) \over |V_{ub}|^2}
&=& ~~8.3 \times 10^{12}/sec~,~~~(\cite{isgw}) \nonumber\\
&=& 32.9 \times 10^{12}/sec~,~~~(\cite{KS}) \nonumber\\
&=& 18.7 \times 10^{12}/sec~.~~~(\cite{WSB}) \nonumber
%\label{eq1}
\end{eqnarray} 

Recently it has been  suggested  that the measurements of
hadronic invariant mass spectrum \cite{kim-ko} as well as 
hadronic energy spectrum \cite{bouzas}
in the inclusive $B \rightarrow X_{c(u)} l \nu$ decays can be
useful in extracting $|V_{ub}|$ with better theoretical understandings.
In future asymmetric $B$-factories with microvertex detector, 
the hadronic invariant mass spectrum will offer
alternative ways to select $b \rightarrow u$ transitions that are much more
efficient than selecting the upper end of the lepton energy spectrum, with
much less theoretical uncertainties.
The measurement of ratio $|V_{ub}/V_{ts}|$ from the differential
decay widths of the processes $B \rightarrow \rho l \nu$ and
$B \rightarrow K^* l \bar l$ by using $SU(3)$-flavor symmetry and
the heavy quark symmetry has been also proposed \cite{sanda}.
It is urgently important that all the available methods of determinating 
$V_{ub}$ have to be thoroughly explored to measure the most important CKM 
matrix element as accurately as possible in the forthcoming $B$-factories. 

\noindent
{\bf 2. Theoretical Discussions}~~~
Over the past few years, a great progress has been achieved 
in our understanding of {\it inclusive} semileptonic decays 
of heavy mesons \cite{hqet}, especially in the lepton energy spectrum.
However, it turns out that the end point region of the lepton energy 
spectrum cannot be described by 
$1/m_{_Q}$ expansion. Rather, a partial resummation of  $1/m_{_Q}$ 
expansion is required \cite{resum}, closely analogous to the leading
twist contribution in deep inelastic scattering, which brings about 
possible model dependences.

Even with a theoretical breakdown near around the end point region of lepton
energy spectrum, accurate predictions of the {\bf total} 
integrated semileptonic decay rate have been obtained \cite{hqet} 
including the first non-trivial non-perturbative
corrections as well as radiative perturbative QCD correction
\cite{kuhn}. The related uncertainties in calculation of the integrated
decay rate have been also analyzed \cite{luke,shifman,ball}.
The total inclusive semileptonic decay rate for $B \rightarrow X_q l \nu$ is
given \cite{shifman} as
\begin{eqnarray}
&\Gamma&(B \rightarrow X_q l \nu) = 
{G_F^2 m_b^5 \over 192 \pi^3} |V_{qb}|^2 \nonumber\\
&\Biggl\{& 
\Biggl[ z_0(x_q \equiv m_q/m_b) - {2 \alpha_s(m_b^2) \over 3 \pi}
g(x_q) \Biggr] \left( 1 - {\mu_\pi^2 - \mu_G^2 \over 2 m_b^2} \right) 
\nonumber\\
&-& z_1( x_q) {\mu_G^2 \over m_b^2} + 
{\cal O}(\alpha_s^2,\alpha_s/m_b^2,1/m_b^3)
~~\Biggr\}~~,
\label{eq2}
\end{eqnarray}
where
\begin{eqnarray}
z_0(x) &=& 1 -8x^2 +8x^6 -x^8 -24x^4\log{x}~~, \nonumber\\
z_1(x) &=& (1-x^2)^4~~, \nonumber\\
{\rm and}~~ g(x) &=& (\pi^2-31/4)(1-x)^2+3/2 \nonumber
\end{eqnarray}
is the corresponding single gluon
exchange perturbative QCD correction \cite{kuhn,kim-martin}.
The expectation value of energy due to the chromomagnetic hyperfine 
interaction, $\mu_G$, can be related to the $B^* - B$ mass difference
\begin{equation}
\mu_G^2 = {3 \over 4} (M_{B^*}^2 - M_B^2) 
\approx (0.350 \pm 0.005)~{\rm GeV}^2~~,
\label{eq4}
\end{equation}
and the expectation value of kinetic energy of $b$-quark inside $B$ meson, 
$\mu_\pi^2$, is given from the various  
arguments \cite{mu-pi,kim-namgung,gremm},
\begin{equation}
0.30~{\rm GeV}^2 \leq \mu_\pi^2 \leq  0.65~{\rm GeV}^2~~,
\label{eq5}
\end{equation}
which shows much larger uncertainties compared to $\mu_G^2$.
The value of $|V_{cb}|$ has been estimated \cite{luke,shifman,ball}
from Eq. (\ref{eq2}) of the total decay rate 
$\Gamma(B \rightarrow X_c l \nu)$ by using the pole mass of 
$m_b$ and a mass difference $(m_b - m_c)$ based on the HQET. 
As can be easily seen from Eq. (\ref{eq2}), the factor
$m_b^5$, which appears in the semileptonic decay rate, 
but not in the branching fraction, 
is the largest source of the uncertainty, resulting in about
$5 \sim 20\%$ error in the prediction of $|V_{cb}|$ 
via the semileptonic branching fraction and $B$ meson life time.
Historically, the ACCMM model \cite{accmm} was motivated to avoid this
factor $m_b^5$, and at the same time to incorporate the bound state effect of 
initial $B$ meson.

The ratio of CKM matrix elements $|V_{ub}/V_{cb}|$  can be determined in 
a model-independent way by taking the ratio of semileptonic decay widths
$\Gamma(B \rightarrow X_u l \nu)/\Gamma(B \rightarrow X_c l \nu)$.
As can be seen from Eq. (\ref{eq2}), 
this ratio is theoretically described by the 
phase space factor and the well-known perturbative QCD correction only,
\begin{eqnarray}
{\Gamma(B \rightarrow X_u l \nu) \over \Gamma(B \rightarrow X_c l \nu)}
= \left| { V_{ub} \over V_{cb} } \right|^2 
&\Biggl[& 1 - {2 \alpha_s \over 3 \pi}
\left( \pi^2 - {25 \over 4} \right) \Biggr] \nonumber\\
\times &\Biggl[& z_0(x_c) - {2 \alpha_s \over 3 \pi} g(x_c) \Biggr]^{-1}.
\label{eq7}
\end{eqnarray}
We strongly emphasize here 
that the sources of the main theoretical uncertainties, 
the most unruly factor $m_b^5$ and the still-problematic non-perturbative 
contributions, are all
canceled out in this ratio. By taking $\alpha_s(m_b^2) = (0.24 \pm 0.02)$,
and using the mass difference relation from the HQET \cite{mass}, 
which gives
$x_c \equiv m_c/m_b \approx 0.25 - 0.30$, 
the ratio of the semileptinic decay widths is estimated as
\begin{equation}
{\Gamma(B \rightarrow X_u l \nu) \over \Gamma(B \rightarrow X_c l \nu)}
\simeq (1.83 \pm 0.28) \times \left| { V_{ub} \over V_{cb} } \right|^2,
\label{eqq8} 
\end{equation}
and the ratio of CKM elements is 
\begin{eqnarray}
\left| { V_{ub} \over V_{cb} } \right| &\simeq&
(0.74 \pm 0.06) \times
\left[ {{\cal B}(B \rightarrow X_u l \nu) \over 
{\cal B}(B \rightarrow X_c l \nu) } \right]^{1/2}, \nonumber\\
&\simeq&
(0.75 \pm 0.06) \times
\left[ {{\cal B}(B \rightarrow X_u l \nu) \over 
{\cal B}(B \rightarrow X l \nu) } \right]^{1/2},
\label{eq8}
\end{eqnarray}
where in the last relation we have assumed ${\cal B}(B \rightarrow X l \nu)
\sim (1.02) \cdot {\cal B}(B \rightarrow X_c l \nu)$.
Once the ratio of semileptonic decay widths (or equivalently the ratio of
branching fractions 
${\cal B}(B \rightarrow X_u l \nu)/{\cal B}(B \rightarrow X_c l \nu)$)
is measured in the forthcoming asymmetric $B$-factories,
this will give a powerful model-independent determination of $|V_{ub}/V_{cb}|$.
We will discuss on this
experimental possibility in details in the next Section.
There is absolutely no model dependence in these ratios 
Eqs. (\ref{eq7},\ref{eqq8},\ref{eq8}).
As explained earlier, for example in ACCMM model \cite{accmm} the model 
dependence comes in via the introduction of parameter $p_{_F}$ to avoid
the factor $m_b^5$, which is now canceled in these ratios.
The problem of the semileptonic branching
fraction ({\it so-called} the discrepancy between the theoretical prediction 
and the actual experimental measurement of 
the semileptonic branching fraction ${\cal B}_{sl}$) would be also
canceled out in the ratio of the branching fractions.

\noindent
{\bf 3. Experimental Discussions}~~~
In order to measure $|V_{ub}/V_{cb}|$ (and $|V_{ub}|$) model-independently
by using the relations Eq. (\ref{eq8}), it is critically
required to separate the $b \rightarrow u$ semileptonic decays 
from the dominant $b \rightarrow c$ semileptonic decays, 
and to precisely measure branching fraction
${\cal B}(B \rightarrow X_u l \nu)$ or the ratio
${\cal B}(B \rightarrow X_u l \nu)/{\cal B}(B \rightarrow X_c l \nu)$.
At presently existing symmetric $B$-factories, ARGUS and CLEO, where $B$ and 
$\bar B$ are produced almost at rest, this required separation is possible only
in the very end point region of lepton energy spectrum, because both $B$ and 
$\bar B$ decay into whole $4 \pi$ solid angle from the almost same decay point, 
and it is not possible for the produced particles to be identified
from which is the original $B$ meson.
However, in the forthcoming asymmetric $B$-factories with microvertex 
detectors, BABAR and BELLE \cite{B-fact}, where the two
beams have different energies and the produced $\Upsilon(4S)$ is not at
rest in the laboratory frame, the bottom decay vertices too will be
identifiable with still greater advantage to the analyses.
The efficiency for the whole event reconstruction could be relatively high
(maybe $1 \sim 10 \%$ efficiency) limited only by about 
$60 \%$ of $\pi^0$-reconstruction efficiency, 
and this $b \rightarrow u$ separation would be experimentally viable option.

As of the most straightforward separation method, the measurements of inclusive 
hadronic invariant mass ($m_{_X}$) distributions in 
$B \rightarrow X_{c,u} l \nu$   can be very useful 
for the fully reconstructed semileptonic decay events.  
For $b \rightarrow c$ decays, one necessarily has  
$m_{_X} \geq m_{_D} = 1.86$ GeV.  
Therefore, if we impose a condition $m_{_X} < m_{_D}$, 
the resulting events come only from $b \rightarrow u$ decays,  
and about 90\% of the $b\rightarrow u$ events survive this cut.   
This is already in sharp contrast with the usual cut 
on charged lepton energy $E_l$.
In fact, one can relax the condition  $m_{_X} < m_{_D}$, and extract
the total $b \rightarrow u$ semileptonic decay rate \cite{kim-ko}, 
because the $m_{_X}$ distribution in $b \rightarrow c$ decays is completely 
dominated by contributions of three resonances $D, D^{*} $ and $D^{**}$, 
which are  essentially like $\delta$-functions, 
\begin{equation} 
{d \Gamma \over d m_{_X}} = \Gamma(B\rightarrow R l \nu)~
\delta(m_{_X} - m_{_R})~~,
\label{eq9}
\end{equation}
where the resonance $R = D, D^*$ or $D^{**}$.
In other words, one is allowed to use the $b \rightarrow u$ events 
in the region even above $m_{_X} \geq m_{_D}$, first by excluding small 
regions in $m_{_X}$ around $m_{_X} = m_{_D}, m_{_{D^{*}}}, m_{_{D^{**}}}$,
and then by including the regions again numerically 
in the $m_{_X}$ distribution of $b \rightarrow u$ decay from its value just
around the resonances. 
We note that there is possibly a question of bias.  Some classes of final
states ($e.g.$ those with low multiplicity, few neutrals) may be more
susceptible to a full and unambiguous reconstruction. Hence an
analysis that requires this reconstruction may be biassed. However,
the use of topological information from microvertex detectors
should tend to reduce the bias, since vertex resolvability depends
largely on the proper time of the decay and its orientation relative
to the initial momentum (that are independent of the decay mode).
Also such a bias can be allowed for in the analyses, via a suitable
Monte Carlo modeling.
For more details on this inclusive hadronic invariant mass distribution
$d \Gamma / d m_{_X}$, please see Ref. \cite{kim-ko}.

Even without full reconstructions of final particles, 
one can separate $b \rightarrow u$ decays from $b \rightarrow c$ decays
by using the particle decay properties \cite{pdg}.  Since 
$D^{**} \rightarrow D^* + \pi$ and $D^* \rightarrow D + \pi$, the semileptonic
$b \rightarrow c$ decays always produce at least one final state $D$ meson,
compared to $b \rightarrow u$ decays which produce particles, $\pi,~\rho,~...$
that always decay to one or more $\pi$ mesons at the end. 
Therefore, the $b \rightarrow u$ decay 
separation can be achieved only with the accurate `$D - \pi$' separation in 
particle detectors. There still is a possible non-resonant decay background 
from $B \rightarrow (D + \pi) l \nu$ in using previously explained  
inclusive $m_{_X}$ distribution separation.
However, with this addtional `$D - \pi$' separation condition 
the $b \rightarrow u$ decays
can be safely differentiated from the dominant $b \rightarrow c$ decays.
There is another possible source of background to this `$D - \pi$' 
separation condition from the cascade decay of 
$b \rightarrow c \rightarrow s l \nu$. Recently ARGUS and CLEO \cite{cas-B} 
have separated this cascade decay background
from the signal events to extract the model-independent spectrum of 
$d\Gamma/dE_l(B \rightarrow X_c l \nu)$ for the whole region of electron
energy, by taking care of lepton charge and $B - \bar B$ mixing 
systematically. In future asymmetric $B$-factories with much higher 
statistics, this cascade decay will not be any serious background at all
except for the case with very low energy electron.  

In view of the potential importance of 
${\cal B}(B \rightarrow X_u l \nu )/{\cal B}(B \rightarrow X_c l \nu )$ 
as a new theoretically model-independent probe for measuring
$|V_{ub}|$ and $|V_{ub}/V_{cb}|$,
we would like to urge our experimental colleagues to make sure that this 
$b \rightarrow u$ separation can indeed be observed.

\begin{center}
{\bf Acknowledgements}\\
\end{center}

\noindent
The work  was supported 
in part by the Korean Science and Engineering  Foundation, 
Project No. 951-0207-008-2,
in part by Non-Directed-Research-Fund, Korea Research Foundation 1993,
%in part by the Center for Theoretical Physics, Seoul National University, 
in part by Yonsei University Faculty Research Grant, and
in part by the Basic Science Research Institute Program,
Ministry of Education 1995,  Project No. BSRI-95-2425.
For more details on this talk, please see Ref. \cite{new-kim}.

\end{document}